\documentclass[twocolumn]{aa}
\usepackage{graphicx}
\usepackage{txfonts}
\begin{document}
   \title{Time-dependent ejection velocity model for the outflow 
 of \object{Hen 3--1475}
 \thanks{Based on observations made with the Hubble 
 Space Telescope, obtained from the
 Data Archive at the Space Telescope
 Science Institute, which is operated by the Association of Universities for Research in Astronomy, Inc., under NASA
  contract NAS5-26555 }}


   \author{P.F. Vel\'azquez
          \inst{1}
          \and
          A. Riera\inst{2,3}
	  \and A.C. Raga\inst{1}
          }

   \offprints{P.F. Vel\'azquez}

   \institute{Instituto de Ciencias Nucleares, UNAM, Ciudad Universitaria,
Apdo. Postal 70-543, CP: 04510, M\'exico D.F., M\'exico\\
              \email{pablo@nuclecu.unam.mx, raga@nuclecu.unam.mx}
         \and
             Departament de F\'\i sica i Enginyeria Nuclear, Universitat
Polit\`ecnica de Catalunya, Escola Universit\`aria Polit\`ecnica de Vilanova i La Geltr\'u, Av. 
V\'ictor Balaguer s/n, E-08800 Vilanova i la Geltr\'u, Spain\\
             \email{angels.riera@upc.es} 
          \and Departament d'Astronomia i Meteorologia, Universitat de Barcelona, Av. Diagonal 647, 
E-08028 Barcelona, Spain.
             }
   \date{Received ; accepted}

   \abstract{We present 2D axisymmetric and 3D numerical simulations 
of the proto-planetary nebula \object{Hen 3-1475}, which is 
characterized by a 
remarkably highly collimated optical jet, formed by a string of shock-excited
knots along the axis of the nebula. 
It has recently been suggested that the
kinematical and morphological properties of the \object{Hen 3-1475} jet could be the 
result of an ejection variability of the central source
(Riera et al. 2003). The 
observations suggest a periodic variability of the ejection velocity 
superimposed on a smoothly increasing ejection velocity ramp.
From our numerical simulations, we have obtained intensity maps (for
different optical emission
lines) and position-velocity diagrams, in order to make a direct comparison
with the HST observations of this object.
Our numerical study allows us to conclude that a model of a precessing jet
with a time-dependent ejection velocity, which is propagating into an ISM
previously perturbed by an AGB wind, can succesfully explain both the
morphological and the kinematical characteristics of this
proto-planetary nebula.
   }

   \maketitle
%

\section{Introduction}
Recent images of young PNe obtained with the HST have revealed that almost 
all of them have aspherical morphologies. Also,
a significant fraction show bipolar 
or mutipolar lobes, highly collimated outflows or jets and other complex 
structures such
as loops (see, e.g., Sahai \& Trauger 1998; Garc\'\i a-Lario, Riera 
\& Manchado 1999; Trammell \& Goodrich 2002). Sahai \& Trauger (1998) 
suggested that the outflows (i.e., collimated fast winds, hereafter CFWs) 
acting during the early post-AGB phase may have an important impact in the 
formation of these complex structures. However, the mechanism(s) producing the 
collimated fast winds or jets are not well understood. 

The most promising scenarios for producing CFWs or jets are~:

\noindent (i) magnetized accretion disk and binary stellar system
(see, e.g., Soker \& Livio 1994; Blackman et al. 2001),

\noindent (ii) a rapidly rotating star with a toroidal
 magnetic field (see, e.g., Garc\'\i a-Segura et al. 1999; Garc\'\i a-Segura 
\& L\'opez 2000).

Recently, the interaction of CFWs and the AGB envelope has been explored
 both analytically and numerically (Soker \& Rappaport 2000; 
 Lee \& Sahai 2003; Garc\'\i a-Arredondo \& Frank 2003). 

The proto-planetary nebula (PPN) \object{Hen 3-1475} shows a very well 
collimated
jet, formed by  a string of knots distributed along the 
main axis of the nebula (reminiscent of a HH object). \object{Hen 3-1475}
 has been the 
object of optical imaging from the ground (Riera et al. 1995) and 
with the Hubble Space Telescope (HST) 
(Bobrowsky et al. 1995; Borkowski, Blondin \& Harrington 1997; 
Ueta, Meixner \& Bobrowsky  2000),
 near-infrared photometry (Garc\'\i a-Lario et al. 1997; 
Rodrigues et al. 2003), optical polarization (Rodrigues et al. 2003),  
optical spectroscopy (Riera et al. 1995;  Bobrowsky et al. 1995; 
Borkowski \& Harrington 2001; S\'anchez Contreras \& Sahai 2001; 
Riera et al. 2002, 2003), and proper motion studies 
(Borkowski \& Harrington 2001; Riera et al. 2002, 2003). 

The central region is probably a dusty disk (or torus). The densest 
parts of the disk are seen in the WFPC2 images of  Borkowski et al. 
(1997) as dark patches. The intrinsic polarization and large  
reddening of the circumstellar region of \object{Hen 3-1475} has been studied 
in detail by Rodrigues et al. (2003).   
The presence of CO molecular emission suggests the existence of 
large amounts of neutral gas. The CO emission shows an expanding torus 
of material (Knapp et al. 1995, Bujarrabal et al. 2001).  
The recent CO observations of Huggins et al. (2004) show the presence of 
an expanding biconical envelope, and provide strong evidence for entrainment of 
the molecular gas by the high-velocity jets. 
The OH emission suggests the existence of an aspherical shell 
expanding at 25 km s$^{-1}$ (Bobrowsky et al. 1995, Zijlstra et al. 2001). 

The radio continuum indicates the presence of a compact, ionized 
region surrounding the central B star (Bobrowsky et al. 1995; 
Knapp et al. 1995). Optical, ground based spectroscopy of the central 
source revelead the presence of  P-Cygni Balmer line profiles 
(Riera et al. 1995), which indicates the presence of
mass outflows at high velocities. 
An analysis of a STIS, long-slit H$\alpha$ spectrum of the central 
region of \object{Hen 3-1475} (S\'anchez Contreras \& Sahai 2001), 
suggests the presence of two different winds 
outflowing from the central star at high velocities. 
The highest velocity outflowing wind is highly collimated close
 to the central star, and shows a radially increasing velocity. 

The emitting knots show extremely broad, double-peaked emission 
line profiles (with FWHM from 400 to 1000 km s$^{-1}$, 
Borkowski \& Harrington 2001; Riera et al. 2003). The radial 
velocities observed along the {\object{Hen 3-1475}} jet decrease with 
increasing 
distance to the central source, showing abrupt velocity changes 
in the transition regions from each (sub)condensation to the next one. 

From a comparison of the spectra observed for the
intermediate knots of \object{Hen 3-1475} with the spectra predicted
by plane-parallel shocks, Riera et al. (1995) concluded that
the emission  is produced in a shock wave which propagates 
through a nitrogen-enriched medium. Predictions from
plane-parallel shock models 
(adopting Type I PN abundances) with shock velocities from 100 
$\rightarrow$ 150 km s$^{-1}$ qualitatively reproduce the observed spectra 
(Riera et al. 2003).  

Recently, {\it Chandra} has detected X-ray emission from \object{Hen 3-1475}
(Guerrero et al. 2004; Sahai et al. 2004). The X-ray source position coincides 
with the innermost optical knot NW1, and 
the X-ray emitting gas is probably shock heated.
The X-ray spectrum implies a temperature of $\sim$ 2.4 $\times$ 10$^{6}$ K, 
and the observed X-ray luminosity  
implies a shock velocity $\sim$ 400 km s$^{-1}$ (Guerrero et al. 2004; 
Sahai et al. 2004).  
     
Recently, it has been suggested  that kinematic and morphological 
properties of the \object{Hen 3-1475} jet could be the result of a variability 
of the central source (Riera et al. 2003). 
During the past 15 years, time-dependent ejection velocity models
have been studied and applied to HH jets, and these models in principle
can also be applied to the \object{Hen 3-1475} jet. In these models, the 
chain of
knots observed in the jet is assumed to correspond to the successive  
internal working surfaces{\footnote{A ``working surface'' is a double 
shock structure produced by the propagation of a fast flow into a
slower flow. This working surface is
formed by a bow shock sweeping up the material of the slow flow, an inward 
shock or Mach disk that decelerates the fast flow  gas, and a 
contact discontinuity
between them, separating the shocked slow and fast flow gas}} 
resulting from a (possibly periodic)
ejection velocity variability. 
One-dimensional analytic (Raga et al. 1990; Cant\'o, Raga \&
D'Alessio 2000), as well as axisymmetric and 3-D 
numerical simulations (V\"olker et al. 1999; Raga et al. 2002;
Masciadri et al. 2002) have been carried out for
different functional forms for the ejection velocity variability. 

In the present paper, we investigate whether or not the properties 
of the jets of \object{Hen 3-1475}
can be the result of the variability of the ejection velocity.
As far as we are aware, our work
is the first attempt to obtain a gasdynamical simulation of this object. 
The origin of both the collimation mechanism and of the
ejection velocity variability is not studied in this paper.

The present paper is organized as follows~: in section 2, a variable
velocity ejection law is obtained for the case of \object{Hen 3--1475}; 
the initial
conditions and assumptions for the 2--D and 3--D numerical simulations are
presented in section 3; in section 4, we present the results of the temporal
evolution from 2--D axisymmetric numerical simulation (subsection 4.1) 
and the comparison of 3--D results with HST observations (subsection 4.2);
finally, in section 5, we give our conclusions.


\section{A variable velocity ejection law for  Hen 3--1475}

We now consider the 
ground-based and the STIS HST spectroscopy, together with
proper motion measurements (taken from Riera et al. 2003) of the
\object{Hen 3-1475} jet. We first deproject
the distance $z$ from the source and the velocity $v(z)$ for an assumed
 angle $\phi=50^{\circ}$ between the outflow axis and
the plane of the sky (Borkowski \& Harrington 
2001). In this way, we obtain the physical velocity of the emitting
gas as a function of distance from the source (see Figure \ref{fig1}).

From Figure \ref{fig1}, we see that the flow velocity decreases in an 
approximately monotonic way as a function of increasing distances from the 
source. Such a velocity configuration suggests a time-dependent ejection 
velocity which smoothly increases as a function of time (with larger velocities
at more recent times). Also, we need to superpose
(on this ``smooth ramp'' variability) a variation in the ejection velocity
with an appropriate period and amplitude in order to obtain internal
working surfaces at positions that approximately correspond to the
positions of the knots along the observed jet.

After trying a number of different functional forms for the ejection
velocity variability, we have chosen the form~:

\begin{equation}
v_j = v_0 + v_1\, sin\, (2 \,\pi  (t-t_0)\,/\,\tau) + a\, t ,
\label{vj}
\end{equation}
\noindent which includes a periodic term (described by a sinusoidal function
 characterized by an amplitude velocity $v_1$ and a period $\tau$), 
superimposed on a linear increase of the velocity (at a rate $a$). 
In Eq. (1), $t_0\,(<0)$  is the time at which the source is initially
turned on  ($t = 0$ corresponding to
the time at which the observations were obtained). Through the computation
of several models with different parameters, we find that
the following parameter set~:
$v_0$ = 400 km s$^{-1}$ (mean velocity),
$v_1$= 150 km s$^{-1}$ (velocity variability half-amplitude),
$\tau$ = 120 yr (period of the sinusoidal veriability), 
$a$ = 1 km s$^{-1}$ yr$^{-1}$ (slope of linear ramp) and
$t_0=-640$~yr (time at which the jet is turned on) 
produces jets with knot structures and kinematical properties
similar to the ones of the \object{Hen 3-1475} jet. We then use the ejection
velocity variability given by equation (1) with this particular set
of parameters in order to compute the axisymmetric and 3D jet simulations
discussed in the following sections.

 \begin{figure}
   \centering
   \caption{Velocity of the emiting gas as a function of distance $z$ from the
source, reconstruted from the ground-based radial velocities (solid circles) 
and the STIS spectroscopy (open circles)}
              \label{fig1}
    \end{figure}

\section{Numerical Simulations}
\subsection{2--D Numerical simulation}

The 2--D numerical simulations have been carried out with
the axisymmetric version of the Yguaz\'u--a code (which is described
in detail by Raga et al. 2000).
This code integrates the cylindrical gasdynamic equations (employing
the ``flux vector splitting'' scheme of van Leer 1982) together with
a system of rate equations for atomic/ionic species.
With these rate equations, a non-equilibrium cooling function is computed.
The reaction and cooling rates are described in detail by \cite{raga02}. 
We have used the mean Type I PN abundances from Kingsburgh \& Barlow (1994). 

The axisymmetric numerical simulations were computed on a 4-level binary 
adaptive grid, with a maximum resolution of 1.172$\times$10$^{15}$~cm along
both axes. The computational domain is of 
(12.0, 1.5)$\times$~10$^{17}$~cm, along the axial ($z$) and
radial ($r$) axes, respectively.

The jet is injected at $z=0$ with the time-dependent
ejection velocity  given by 
Eq.(\ref{vj}) (and with the paremeter set given in \S 2).  
The jet has an initial radius r$_j$ = $6\times 10^{15}$~cm, 
(time-independent) density of 500~cm$^{-3}$ (similar to the density
derived by Riera et al. in preparation) and temperature T$_j$ = 1000 K.

We have carried out 2--D numerical simulations with two 
density distributions for the surrounding medium.    
First, we have assumed that the surrounding environment 
is homogeneous, with a density n$_{env}$ = 10~cm$^{-3}$ and
temperature T$_{env}$ = 100 K.

However, the high-velocity outflows of Hen 3-1475 are moving through 
the slow wind previously ejected by the star as a red giant on the 
AGB. The circumstellar envelope of Hen 3-1475 is far 
from being spherical as shown from the optical polarization
maps and the spatial distribution of the  CO and OH 
emission (see, e.g., Bobrowsky et al. 1995, Zijlstra et al. 2001, 
Bujarrabal et al. 2001, Rodrigues et al. 2003, Huggins et al. 2004). 
In order to include this, we have computed a second model, in which
the outflow is moving into an AGB wind. For this model, 
we have assumed that the AGB remnant has an aspherical  
density distribution with high densities on the equator and lower densities 
on the poles, which is described by the following equation (taken from 
Mellema 1995), 
\begin{equation}
\rho(R)=\rho_0 \biggl[1-\alpha \bigg({{1-exp(-2 \beta sin^2\theta)}
\over{1-exp(-2\beta)}}\biggr)\biggr] (R_0/R)^{2}
\label{agb}
\end{equation} 
\noindent where $R=\sqrt{r^2+z^2}$ and $\theta$ is the angle with
respect to $r-$axis. The parameter $\alpha$ determines the ratio between 
the density at the equator and at the pole, and $\beta$ determines the 
way the density varies from the equator to the pole (see Mellema 1995).
The value of $\rho_0$ is calculated from 
the mass loss rate as~: $\rho_0$ = $\dot{M}$ / 4
$\pi$ $R_0^2$ v$_{env}$, where v$_{env}$ is the expansion velocity 
of the AGB remnant. We have chosen $R_0$ = r$_j$, 
a mass loss rate of 10$^{-6}$   
M$_{\sun}$ yr$^{-1}$ and an expansion velocity of 20 km s$^{-1}$ .  
For $\alpha$ and $\beta$ we adopt values of 0.7 and
3.0, respectively. With these parameters and Eq.(\ref{agb}) 
the range of the
AGB wind density along $z$-azix is $\sim$~[500, 0.05] cm$^{-3}$ from the 
injection point to the $z$ boundary, respectively.

\subsection{3--D Numerical simulation}
Axisymmetric numerical simulations are powerful tools for finding a
time-dependent ejection velocity (such as the one given by Eq.(\ref{vj}))
which produces knots with appropriate positions and kinematical
properties. However, the \object{Hen 3-1475} jet
has a ``S''-like morphology, which is far from axisymmetric.
Then, in order to reproduce this ``S''-like shape, we have carried out
3--D numerical simulations, including a precession of the outflow
axis with a period of 1500~yr and a half aperture angle for the precession
cone of 7.5$^{\circ}$.
This precession produces approximately one half ``wave'' of the approximately
helical jet beam within the 
observed length of the jet in \object{Hen 3-1475} (in agreement with the 
observed images).

We then carry out a 3-D numerical simulation of a jet with the
ejection velocity variability 
given in Section 2, and the precession described above. 
The numerical simulations were carried out with the 3--D version
of Yguaz\'u--a code, employing a five-level binary adaptive grid with a maximum
resolution of 2.34$\times 10^{15}$~cm. The computational domain was
of (3, 3, 12)$\times 10^{17}$~cm along $x-$, $y-$ and $z-$~axes, respectively.

The outflow is injected at $x=y=1.5\times 10^{17}$~cm, on the $z=0$ plane.
The initial jet radius is 1.5$\times$10$^{16}$~cm, the
initial number density is $n_j$ = 
500 cm$^{-3}$ and the temperature is T$_j$ = 1000 K. We have adopted
a larger initial jet radius than in the axisymmetric simulations (see
\S 3.1) in order to resolve the jet beam with $\sim 10$ grid points
(in the lower resolution, 3-D simulation).

In the 3-D simulation, we assume that the jet is moving into
the AGB wind, for which we adopt the aspherical density function 
and parameters described in Section 3.1. 

\section{Results}
\subsection{Temporal evolution of jet structure for 2D numerical simulations}
   \begin{figure}
   \centering
   \caption{Time evolution of the density stratification for the case of
an outflow moving into a homogeneous environment. Integration
times range from 100 to 600 years (top to bottom panels, 
respectively). The logarithmic greyscale, in units of g~cm$^{-3}$,
is given by the vertical bar on the right of the bottom panel.
The scale in both axes is given in units of 10$^{17}$~cm.}
              \label{fig2}
    \end{figure}

  \begin{figure}
   \centering
   \caption{Same as Fig.\ref{fig2} but for the case of a jet expanding into
an AGB remnant with a density distribution given by Eq.\ref{agb} (see text)}
              \label{fig3}%
    \end{figure}

Figure \ref{fig2} shows a 600~yr time-sequence of the density 
stratification (in the $rz$-plane) for the case of a jet moving into a
homogeneous environment. Figure \ref{fig3}
shows a similar time-sequence, but for
a jet propagating into the AGB wind (with a density 
distribution given by Eq.(\ref{agb})).

The gray-scale representations of the density stratifications obtained for 
integration times from 100 to 600 years (see Fig.\ref{fig2} and \ref{fig3}) show the 
head of the jet traveling away from the injection point. 
The ejection velocity time variability produces
internal working surfaces that travel down the jet (which we identify with the
emitting knots of the observed jet). 
Each working surface is produced when fast material catches up 
with slower material ejected at earlier times.

The temporal variability law used for the velocity in these simulations 
corresponds to the
superposition of two terms~: a periodic
variability (described by a sinusoidal function) and a linear increase
in the velocity. While a linearly increasing ejection velocity vs. time
produces a single, accelerating leading working surface (see Raga et al.
1990 and Cant\'o, Raga \& D'Alessio 2000), the sinusoidal mode (which is
superimposed on the linear ejection velocity vs. time dependence, see Eq. 1)
produces a train of internal working surfaces which travel down the jet beam
(see, e.~g., Raga \& Noriega-Crespo 1998).

A comparison between Fig.\ref{fig2} and Fig.\ref{fig3} illustrates the effect of the 
stratified environment on the evolution of the knots. Initially (at times
$<300$~yr), the head of the jet propagates faster in the uniform environment
model. However, as the time-evolution proceeds, the head of the jet
in the AGB wind case accelerates as it gets out to the lower density
regions farther away from the source, and the jet eventually becomes
longer than the one of the homogeneous environment jet model (see the
$t=600$~yr model of Figs. \ref{fig2} and \ref{fig3}).

  \begin{figure}
   \centering
   \caption{Comparison between simulated H$\alpha$ emission for a model
with a jet propagating into an uniform ISM (bottom panel), and  the 
corresponding emission for the case of a jet moving into an AGB wind 
(top panel). Both maps have been obtained integrating the H$\alpha$ emission 
coefficient along lines of sight, for an angle of $50^{\circ}$ between
the flow axis and the plane of the sky. Both maps correspond to
a time-integration of 640~yr. The
horizontal and vertical axes are given in arcseconds (for a distance
of 5.8 kpc to the object). The H$\alpha$ emission is depicted with a logarithmic
scale (in units of erg~$\rm s^{-1}\ \rm{cm}^{-2}\ \rm{sr}^{-1}$) given by
the vertical bar on the right of the bottom panel.}
              \label{fig4}
    \end{figure}

An interesting difference between the homogeneous environment and the AGB
wind models can be seen in the H$\alpha$ emission maps (Fig.\ref{fig4}).
While the uniform environment jet model shows extended bow shock wings
(associated with the leading working surface), the AGB wind model does
not show such wings. This is a direct result of the fact that at distances
of $\sim 10^{18}$~cm from the source the AGB wind has a very low density,
so that the emission from the bow shock is much lower than in the
homogeneous environment model.

In this way, we see that the model of a jet travelling into an AGB wind
produces H$\alpha$ intensity maps that more closely resemble the observations
of the \object{Hen 3--1475} jet, as this object does not show emission from extended
bow shock wings associated with the jet head. Because of this, we carry
out 3D simulations only for the case of a jet moving into an AGB wind,
and do not study 3D models of jets moving into a uniform environment.

  \begin{figure}
   \centering
   \caption{Density (left), pressure (centre) and temperature (right) 
stratifications for a precessing jet moving into
an AGB wind, obtained from the 3D numerical simulation described
in the text. Each 
panel shows a cut (on the $y-z$ plane) of the corresponding variable for a
640~yr integration time. The scales of both axes are in
units of 10$^{17}$~cm. The density, pressure and temperature scales are
logarithmic (top horizontal bar on each panel), given in g cm$^{-3}$
dyn cm$^{-2}$ and K, respectively.}
              \label{fig5}%
    \end{figure}
  \begin{figure}
   \centering
   \caption{[N~II] 6583 \AA, [S~II] (6717+6730) \AA, [O~I] 6300 \AA,[O~III] 5007 \AA~ and 
H$\alpha$ emission maps (from left to right), corresponding to
an integration time of 640~yr. These maps have been obtained integrating
the emission line coefficient along of lines of sight, considering that
the jet axis has an angle of 50$^{\circ}$ respect to the plane of the sky.
The horizontal and vertical axis are in arcseconds (after considering a
distance of 5.8 kpc) while the vertical bar
on the right, shows (in logarithmic scale) the normalized emission.} 
              \label{fig6}%
    \end{figure}
 \begin{figure}
   \centering
   \caption{HST image in [N II] of Southeastern outflow of 
\object{Hen 3--1475}, which exhibits a jet-like structure and  three pairs of 
symmetric emission knots. The innermost pair of knots show subarcsecond 
structure, with the existence of three well defined compact 
subcondensations. The knots have been labeled following Riera et al. (2003).}
              \label{fig7}%
    \end{figure}

\subsection{Comparison between 3D numerical results and HST observations}

Figure \ref{fig5} shows cuts on the $y-z$ plane of the density (left),
pressure (centre) and temperature (right) stratifications, for the 3D
simulation of a precessing jet moving into an AGB wind (see \S 3.2). 
These stratifications correspond to an integration time of 640~yr.

From the density stratification (left panel of Fig.\ref{fig5}), we see  that 
the jet has a leading head, followed by two high density knots, which
coincide with high-pressure regions (central panel).  
There is a high-temperature ($>10^6$~K) region which corresponds to
the ``cocoon'', while the internal jet regions have temperatures
$\sim 10^5$~K or lower (right hand panel of Fig.\ref{fig5}).

The main effect of the precession is to produce a curved morphology for
the jet beam, which qualitatively resembles the ``S''-shaped morphology of the 
\object{Hen 3--1475} jet (see Fig.\ref{fig5}). Otherwise, the 3D model has
internal working surfaces which are similar to ones of the axisymmetric
models (see \S 4.1).

The predictions obtained from the 3D model can be compared 
with the WFPC2 HST images and with the position-velocity diagrams 
obtained from the (spatial and spectral) high-resolution STIS spectroscopy.  
In order to compare the 3D numerical simulation with the HST 
images of \object{Hen 3--1475}, from the model
we have computed intensity maps for the H$\alpha$, 
[N~II] 6583 \AA, [O~I] 6300 \AA, [S~II] (6717+6731) \AA~ and [O III] 5007 \AA~
(from left to right, Fig. \ref{fig6}) integrating the emission coefficient
along lines of sight. We have  assumed that the 
$y$-axis lies in the plane of the sky, and that the $z$-axis is oriented
toward the observer at an angle of 50$^{\circ}$ with respect to the plane
of the sky. Furthermore, we have changed the
physical coordinates of the model
to angular sizes (after taking into account distance and projection effects).

In Figure \ref{fig6}, we show the simulated line  
emission maps for a 640~yr integration-time,  
which can be compared with the HST 
[N II] image of the SE jet  (shown in Fig. \ref{fig7}).
The images of \object{Hen 3--1475} obtained with the HST  using 
narrow band filters (such as F658N)  
essentially trace the ionized gas through the detection  of the 
nebular emission coming from the corresponding emission line 
(such as [N II] 6583 \AA). However, we should note that losses are expected 
in the detection of the extremely blueshifted and redshifted emission 
due to the large velocities involved. 
Therefore, the observed intensities of the knots in the [N II] image  
are underestimated due to the reduced transmission of the filter F658N.
The losses are larger for the innermost knots, which show the largest 
radial velocities. 

All of the maps (from high to low-ionisation species) show the same 
number of knots (at the same distances from the injection point)  
with similar morphologies. 
Comparing the observed [N II] image (Fig. \ref{fig7}) with the simulated 
emission line maps (Fig.\ref{fig6}), it is clear that the observed and 
simulated maps are qualitatively similar.  

At $t=640$~yr (see Fig. \ref{fig6}), we see the leading head at an angular 
distance  of $\sim$ 7\farcs3 from the injection point, which agrees
well with the  observed angular distance of 7\farcs57 from the central
source to the outermost knot SE3.
In all of the simulated emission line maps we observe that the leading head 
has a double structure. Actually, this double feature corresponds to
two working surfaces. This pair of knots is formed by a faster working
surface which at $t=640$~yr has just overtaken a slower working surface
ejected earlier (this dynamical effect can be seen in the PV diagrams
of Fig. 8).

The intermediate knot (labeled SE2 in Fig. \ref{fig7}) is located at 
a distance of $\sim$ 5\farcs9 in both the observed and simulated images. 
The observed innermost knot (labeled SE1a, SE1b and SE1c) is an elongated 
structure formed by three subcondensations (at distances from the source from 
$\sim$ 2\farcs27 to 3\farcs65). No emission is observed in the innermost 
region (i.e. between SE1a and the central source). In the inner region, the 
simulated maps (see Fig. \ref{fig6}) show an elongated feature connecting the 
point of injection with a compact knot at $\sim$ 4\farcs5 down the jet. 
This feature clearly resembles the morphology observed in the innermost
knot (SE1) of the \object{Hen 3--1475} jet. However, the calculated emission maps
do not show the observed emission gap between this knot and the source.

A detailed comparison of the intensity maps of low-ionization 
(i.e. [O I] or [S II]) and high-ionization species (i.e. [O III]), reveals 
that the leading bow-shock  is prominent in 
[O III] and H$\alpha$, and much fainter than the other knots in the 
low-ionization species (see, e.g.,  the [O I] intensity map of 
Fig \ref{fig6}). 
The simulated [O I] intensity map shows a large range of 
intensities from 
the bright innermost knot to the fainter leading head. All knots show a 
more or less constant [O III] intensity (from the inner knots to the head). 
Unfortunately, no spectrophotometric observations of the three  knots of 
\object{Hen 3--1475} (covering the wavelength range from 5000 to 6800 \AA) have been 
published.   

Figure \ref{fig8} displays position--velocity (PV) diagrams
in [NII] 6583 \AA~ (left) 
and H$\alpha$  (right), obtained from our simulations
for an integration time of 640~yr. Both maps have been 
convolved with a spectral resolution of $50\ {\rm km s}^{-1}$, in order to
simulate the STIS spectral resolution. 
In order to have predictions that
can be directly compared with the spectroscopic 
observations (Borkowski \&  Harrington 2001; Riera et al. 2003;
Riera 2004), the line profiles have been computed assuming a broad 
spectrograph slit (covering all of the computational domain along
the $y$-axis) straddling the $z$-axis. Again, an angle of $50^\circ$
between the $z$-axis and the plane of the sky has been assumed.

The calculated PV diagrams (see Fig. \ref{fig8}) qualitatively reproduce
several characteristics of the observed kinematics. As expected, the large 
radial velocities and the observed decrease of the radial velocity 
with distance along the \object{Hen 3--1475} jet are well reproduced by the model.
Also, the models show distinctive step-like radial velocity decreases 
along the jet axis, and such an effect is also seen in the observed
PV diagrams (see Fig. \ref{fig8} and Riera 2004). 

The observed emission line profiles arising from the knots show wide 
double-peaked profiles (Borkowski \& Harrington 2001; Riera et al. 2003) . 
In the innermost knots the  high-velocity peak is 
stronger than the low-velocity peak, while the situation reverses at the 
intermediate knots.  The calculated PV diagrams show  broad 
emission line profiles, but which 
are narrower than the observed
profiles by a factor of $\sim$ 2. Also, the calculated PV maps
show double-peaked profiles only at very specific spatial locations
(e.~g., in the knot at $\approx$ 4\farcs7 from the source in Fig. \ref{fig8}).

\section{Discussion and Conclusions}

We have carried out 2D (axisymmetric) and 3D numerical simulations for modeling
the jet of the proto-planetary nebula \object{Hen 3--1475}, which is the first attempt 
for obtaining a full and dynamical model for the outflows associated
with this nebula.

We have found that it is possible to fit the observed structure of the
\object{Hen 3--1475} jet with a time-dependent jet model with an ejection velocity
history composed of a sinusoidal mode superimposed on a linear ramp
(i.e., a linear increase of the ejection velocity with time). 
The sinusoidal mode has a period of $\sim$ 120 years and a  
half-amplitude of 150 km s$^{-1}$, and the linear term has a
slope of 1 km s$^{-1}$ yr$^{-1}$. We have explored models of jets
moving into a homogeneous environment, as well as models of jets
moving into a previously ejected, stratified AGB wind.

We should note that
Lee \& Sahai (2003) have presented 2D hydrodynamical simulations of the 
interaction of a CFW/jet and a spherical AGB wind in a PPN. Lee \& Sahai 
adopted a periodic variation in the density and velocity of the jet
(with an amplitude of 150 km s$^{-1}$ and a period of 22 yr, keeping
a constant mass loss rate) in order to reproduce the morphology and
velocities observed in the PPN \object{CRL 618}. A model including both  
periodic density and ejection velocity variations could describe the 
characteristics of PPNs such as \object{Hen 3--1475}. However, Lee \& Sahai (2003) 
 point out some difficulties for producing the knotty structure observed 
along the axis of PPN \object{CRL 618}. It is unclear whether or not this 
kind of 
velocity and density variability would work well for modelling the knot
structure of the PPN \object{Hen 3--1475}.

From the numerical models we have obtained simulated [NII], 
[SII], [OI], [OIII] and H$\alpha$ maps, as well as PV diagrams, which can
be directly compared with the observational results which have
been previously obtained with the HST. 
A comparison between the simulated emission maps (obtained
from the 2D and 3D numerical
simulations) and the observations shows that a model of a
variable speed precessing jet (with
a period of 1500~yr and the ejection velocity history discussed above)
moving into a previously ejected AGB wind produces the best agreement
(of all of the computed models) with the observed characteristics of
the \object{Hen 3--1475} jet.

The observed string of shock-excited knots is well reproduced in our
simulated maps.  Also, the
knot positions are similar to the observed ones, and are distributed
along a curved locus which resembles the S-like shape observed in HST images
of the \object{Hen 3--1475} jet (see Figs. \ref{fig6} and \ref{fig7}).

With respect to the kinematical characteristics, by comparing the
calculated and the observed PV diagrams we can say that our model succesfully
explains~:

\begin{itemize}

\item the large observed radial velocities,

\item the decrease of the radial
velocity of the knots with increasing distances from the source,

\item step-like changes in radial velocity at the positions of the knots,

\item the existence of broad emission line profiles,

\end{itemize}

However, the simulated profiles from 3D simulations are narrower than 
the observed ones (by a
factor of 2), and do not show double-peaked profiles with strong emission
at the highest velocities.

From this discussion, we conclude that many of the morphological and
kinematical features of the \object{Hen 3--1475} jet can indeed be reproduced
(at least in a qualitative way) by a model of a variable ejection velocity,
precessing jet. This quite promising result gives a framework for
modelling and interpreting future, more detailed observations of this jet,
which could lead to more stringent constraints for the proposed model.
Finally, if taken seriously, the parameters of our model give an
idea of what are the variability periods and amplitudes (as well
as the precession) associated with the ejection from \object{Hen 3--1475},
which in principle provide constraints on the nature of the stellar source.

  \begin{figure}
   \centering
   \caption{PV diagram for emission in [NII] (left panel) and H$\alpha$
(right panel), for an integration
time of 640~yr. An angle of 50$^{\circ}$ between the plane of the sky
and the jet direction has been considered. The horizontal axis is given
in units of km s$^{-1}$ while the vertical axis is in arcseconds. 
The [NII] and 
H$\alpha$ emission are shown in logarithmic scale by the bar on the right,
in units of erg s$^{-1}$ cm$^{-2}$ sr$^{-1}$}
              \label{fig8}%
    \end{figure}

\begin{acknowledgements}
Authors thank Saul Rappaport (the referee) for his useful suggestions which
help us to improving the previous version of this manuscript. 
ACR and PFV acknowledge financial support from  Conacyt (M\'exico) grants 
36572-E and 41320-E, and DGAPA (UNAM) grant IN 112602. 
The work of ARi was supported by the MCyT grant AYA2002-00205 (Spain).  
We also acknowledge Israel D\'\i az for maintaining and supporting our
multiprocessor Linux server, where we have carried out our numerical 
simulations, and Antonio Ram\'\i rez for computer help. 

\end{acknowledgements}

\end{document}